\documentclass[a4paper,10pt]{article}
\usepackage[cp1251]{inputenc}
\usepackage[english]{babel}

\usepackage{amssymb}\usepackage{amsmath}
\usepackage{epsfig}
\usepackage{bm}
\usepackage{graphics}
\def\vert{\rule[-2mm]{.1mm}{5mm}}

\def\be{\begin{equation}}
\def\ee{\end{equation}}

\newtheorem{thrm}{\bf Theorem}
\newtheorem{xmpl}{\bf Example}

\newtheorem{rmk}{\bf Remark}

\begin{document}

\title {Real meromorphic differentials:\\ 
a language for the meron configurations in planar nanomagnets}
\author{\copyright 2016 ~~~~A.B.Bogatyrev
\thanks{Supported by RFBR grants 16-01-00568, and RAS Program
"Modern problems of theoretical mathematics"}}
\date{}
\maketitle

Ferromagnets at the nano-scale contain an intriguing world of interacting quasi-particles, the vortex-like patterns of  
magnetization distribution in the material. The concepts of the skyrmions, instantons and merons were invented many years ago in 
high energy physics, however they were adopted also for condensed-matter systems where  vortex-like topologies  describe e.g. 
quantum Hall systems, certain liquid-crystal phases, and Bose-Einstein condensates to name a few. Magnetic vortexes are very important 
in the emerging industry of spintronics where they are considered as the prospective candidates for transporting and storing information 
because they are only a few nanometers in size, very stable, and are easily manipulated with little consumption of energy. 
Taking the industrial applications into account, the flat configurations of magnets become especially important.

The Hamiltonian which governs the magnetization distribution dynamics takes quite many interactions into consideration like the exchange one,
the magnetic dipole one, the Dzyaloshinskii-Moriya, interaction with the external magnetic field, the magnetostatic energy..
which make up a certain hierarchy. There is a consent among the physicists that for the planar ferromagnets at nano-scale, the exchange 
interaction gives the largest input in the behavior. One of the approaches to study magnetization distributions (see \cite{Metlov} and references therin) 
is to minimize the leading Heisenberg exchange term thus getting the metastable states of the magnet whose further dynamics is determined by the next 
in the hierarchy terms of the Hamiltonian.

In this paper we use the language of real meromorphic differentials from the theory of Klein surfaces 
\cite{SS,GrnL} to describe the metastable states of multiply connected planar ferromagnetic nanoelements which minimize the 
exchange energy and have no side magnetic charges. The latter property minimizes part of the magnetostatic energy. Those solutions 
still have enough internal degrees of freedom which may serve as the Ritz parameters for minimization of further relevant 
energy terms or as the dynamical variables for the adiabatic approach. The nontrivial topology of the magnet itself brings us to
several effects first described for the annulus \cite{BM} and observed in the experiment.  Here we explain the topological constraints on the 
numbers of vortexes and antivortexes in the magnet, as well as the algebraic constraints on their positions which stem from 
the Abel's theorem. 

The use of multivalued Prym differentials bring us to new meron configurations which were not considered in the 
seminal work of David Gross \cite{Gr}.

{\bf Acknowledgements}: the author is indebted to Dr. K.L.Metlov from Donetsk Inst. of Physics and Technology 
for the introduction to the magic world of magnetism and for setting  physically significant mathematical problems.

\section{Setting the problem}
\label{Sect1}
In this section we briefly formulate the model for describing magnetization 
textures in planar nanomagnets proposed in 
\cite{Met1, Metlov}. It is based on assuming a well defined hierarchy
among the various energy terms, corresponding to the interaction of local
magnetic moments (such as the exchange and the magnetostatic interactions
between the magnetization vectors). This hierarchy usually holds  in
magnets of sub-micron sizes (nanomagnets). In planar nanomagnet, having a
shape of thin cylinder, the magnetization texture at remanence (no
externally applied magnetic field) can be expressed via a
meromorphic function of complex variable, resulting from solving the
Riemann-Hilbert boundary value problem. This problem can be naturally
reformulated in terms of real meromorphic differentials, which are the main
subject of the present paper. 

The exchange energy in a 2D nanomagnet filling the planar domain $\Omega$ has the appearance \cite{Aha}
\be
\label{ExchE}
E[m]=\int_\Omega \sum\limits_{a=1}^3 |\nabla m_a(x_1,x_2)|^2 d\Omega,
\ee
where (normalized) magnetization distribution $m(x_1,x_2)=(m_1,m_2,m_3)$ is a smooth map from the domain $\Omega$
to the unit sphere: $|m|^2:=\sum_a m_a^2=1$. 

Since both the target and source spaces of the magnetization mapping $m$ are surfaces, there is a temptation to  
use the language of complex variables for this problem. To this end they introduce the complex variable $z:=x_1+ix_2$ varying in the domain $\Omega$ 
and the stereographic projection of the sphere to the complex plane of variable $w$:
\be
\begin{array}{ll}
m_1+im_2=&\displaystyle{\frac{2w}{1+w\bar{w}}},\\
m_3=&\displaystyle{\frac{1-w\bar{w}}{1+w\bar{w}}}.
\end{array}
\ee
The exchange energy can be reformulated in new terms as 
\be\label{ExchEstereo}
E[w]=\int_\Omega \frac8{(1+w\bar{w})^2}(|w_z|^2+|\bar{w}_z|^2)d\Omega,
\ee
where we used the Wirtinger notation for complex differentiation $\partial/\partial z$, 
note that the function $w(z)$ may be neither analytic nor antianalytic.
The Euler-Lagrange equation for the latter functional takes the form:
\be\label{ELeq}
w_{z\bar{z}}=\frac{2\bar{w}}{1+w\bar{w}}w_zw_{\bar{z}}.
\ee

The general solution for this equation is not known (at least to the author). Two wide classes of local solutions are 
respectively instantons $w=f(z)$ introduced by Belavin and Polyakov \cite{BP} and merons (or half-instantons) $w=f(z)/|f(z)|$ 
discovered by Gross \cite{Gr}, where $f$ is a (anti)holomorphic function of $z$. It is easy to show that the general solution of (\ref{ELeq}) with magnetization $m$ in the 
plane of the magnetic is exactly the meronic one.  In the latter case the magnetization  has the appearance 
$m(z)=(\cos(\Phi), \sin(\Phi),0)$ and the exchange energy (\ref{ExchE}) becomes a Dirichlet integral for the real valued function 
$\Phi(z)$:
$$
E[m]=\int_\Omega (\sin^2(\Phi)+\cos^2(\Phi))|\nabla \Phi|^2 d\Omega.
$$
This means that $\Phi(z)$ is harmonic, and locally it is an imaginary part of some holomorphic function $h(z)$.
Thus we get a (local)  meronic representation for  $w(z)$ with $f=\exp(h(z))$.

So, given any meromorphic function $f$ in the domain one can construct either an instanton or a meron.
Note that merons corresponding to $f(z)$ with zeros or poles have an infinite energy (\ref{ExchEstereo}). In this respect D.Gross wrote \cite{Gr}:
"infinite action configurations may be of physical relevance if their action only diverges logarithmically with the volume``.
In our case the energy diverges as the logarithm of the volume of zeros and poles small vicinities. Given a meromorphic function $f$,
they also consider \cite{Gr,Metlov} the finite energy mixture of two basic solutions like 
$$
w(z)=\left\{
\begin{array}{ll}
 f(z)/E_1,& |f(z)|\le E_1,\\
 f(z)/|f(z)|, & E_1\le |f(z)|\le E_2,\\
 f(z)/E_2, &E_2\le|f(z)|,\\
\end{array}
\right.
\qquad ~~ or ~~~~
w(z)=\left\{
\begin{array}{ll}
 f(z)/E_1,& |f(z)|\le E_1,\\
 f(z)/|f(z)|, & E_1\le |f(z)|\le E_2,\\
 E_2/\overline{f(z)}, &E_2\le|f(z)|.\\
\end{array}
\right.
$$
The latter function is continuous however not smooth, the normal derivative breaks at the lines 
$z:\quad |f(z)|=E_1,~ |f(z)|=E_2$ and the equation (\ref{ELeq}) evaluated at those composite solutions acquires the forcing concentrated on the lines.

To obtain a physically meaningful global solution we have to equip it with the relevant boundary condition. 
Without affecting the equation (\ref{ELeq}), they impose an impermeability condition which emulates the magnetic domain wall on the boundary: 
$w(z)$ is (almost everywhere) parallel to the boundary \cite{Metlov,BM1,BM}. 
This choice allows us to get rid of the side magnetic charges \cite{Aha} and therefore of their magnetostatic energy.
Note that critical points of the energy functional (\ref{ExchEstereo}) require different boundary condition, namely the Neumann one:
$w_zdz-w_{\bar{z}}d\bar{z}=0$ at the boundary.

\begin{figure}[h!]
\includegraphics[scale=.85, trim = 1.5cm 24cm .5cm 1cm, clip]{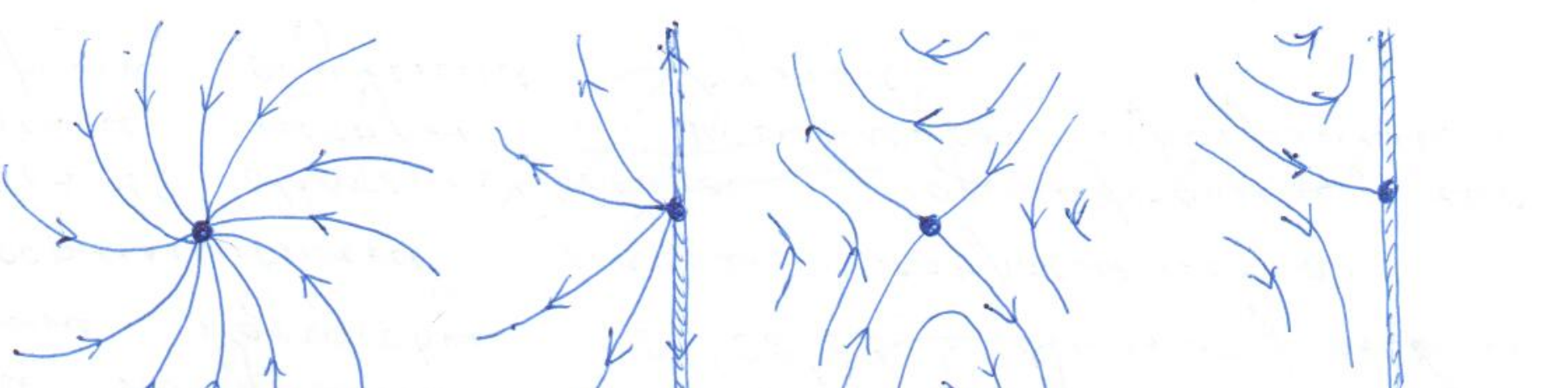}
\caption{Simple (anti) vortexes inside the domain and on its boundary} 
\label{Vortex}
\end{figure}

Summarizing, we extract a reasonable  mathematical problem setting:
{\sf find meromorphic in the domain $\Omega$ functions $f$ such that 
the differential $d\xi:=dz/f(z)$ is real on the boundary of the domain.}

This is exactly the problem of describing all \emph{real meromorphic differentials} in the domain.
The transition from functions $f$ to the differentials $d\xi$ reveals the conformal invariance of the initial problem.
Indeed, the conformal mapping transfers real differentials from one domain to the other (with the same conformal moduli) 
thus keeping not only solutions of the differential equation but also the impermeability boundary condition. 
Due to Helmholz, real meromorphic differentials also describe the stationary motions (with possible sources or sinks) of the ideal fluid in a planar domain
thus giving yet another link between the subjects of hydrodynamics and (micro)magnetics.

\section{Schottky Double of the domain and real differentials} 
Let $\Omega$ be a flat domain of finite connectivity with smooth boundaries.
It's Schottky double is a closed Riemann surface of genus equal to the above mentioned connectivity minus 1 and obtained as follows \cite{SS}:
take two copies of the domain and attach their boundary components one to the other in a natural manner.
Let $z$ be a (local) complex coordinate on one copy of $\Omega$, then $\bar{z}$ will be the (local) complex coordinate on the other copy.
We denote thus obtained surface as ${\cal D}\Omega$, it admits an \emph{anticonformal involution} (also called \emph{reflection}) 
$\tau$ consisting in the interchange of two copies of the domain. The fixed points of this reflection are exactly 
the boundaries of the original domain, called \emph{real ovals} of the surface in this setting. 
The reflection $\tau$ naturally acts on the closed contours on the surface thus splitting the 
1-homology classes into even (=surviving under $\tau$) and odd (=changing their sign) \cite{B2,SS}.
Another natural action of the involution is that on the space of holomorphic (meromorphic) differentials:
\emph{variable change caused by the reflection with subsequent complex conjugation}. The latter action splits the differentials into real -- 
remaining intact -- and imaginary -- changing their sign under this action \cite{SS,GrnL}. Real differentials are exactly those whose restriction 
to any real oval is real. One easily checks that the integration of the real/imaginary differentials over even cycles is a real/imaginary number
while pairing with the odd cycles is respectively imaginary/real \cite{B2}.

The description of all real differentials is in a sense trivial:
take any meromorphic differential on the double ${\cal D}\Omega$ and symmetrize it 
with respect to the induced action of the reflection $\tau$.

\subsection{Examples}
\label {Examples}
\begin{xmpl} 
Real meromorphic differentials in a doubly connected domain may be enumerated in this way.
Any topological annulus is conformally equivalent to the extended complex plane of variable $z$ with two removed  symmetric real slots, say 
$\pm[1,1/k]$,  $0<k<1$ being the modulus of the domain. The double of this domain is equivalent to the elliptic curve 
$$
(z,w)\in\mathbb{C}^2:\qquad
w^2=(z^2-1)(k^2z^2-1)
$$
with anticonformal involution $\tau(z,w)=(\bar{z}, -\bar{w})$. A meromorphic differential on this surface 
has the appearance 
\be\label{RatDiff}
d\omega(z,w)=(R_1(z)+wR_2(z))dz
\ee
with rational functions $R_1$, $R_2$  of the variable $z$. Among them real differentials are exactly those 
satisfying $d\omega(\bar{z},-\bar{w})=\overline{d\omega(z,w)}$, that is with real rational functions $R_1(z)$ and $iR_2(z)$. 
\end{xmpl}

\begin{xmpl} 
Three-connected domains may be treated in a similar way: any pair of pants is conformally equivalent to 
the complex plane with three removed disjoint real slots, say $\cup_{j=1}^3[e_{2j-1},e_{2j}]$,
where three of the points $e_1<e_2<\dots<e_6$ may be normalized at will, three other serve as conformal moduli of pants.
The double of this domain is the genus 2 complex curve
\be
\label{PantsDble}
(z,w)\in\mathbb{C}^2:\qquad
w^2=\prod_{j=1}^6(z-e_j),
\ee
with the reflection $\tau$ as in the previous example. Again, real differentials have the appearance
(\ref{RatDiff}) with real rational functions $R_1(z)$ and $iR_2(z)$.
\label{Pants}
\end{xmpl}

\begin{xmpl}
Real differentials may be obtained in terms of theta functions as well. An annulus  $1\le|z|\le R$
we represent as a factor of the vertical strip $0\le Re(u)\le 1/2$ by a group of translations generated by 
$iT$, where $T=\pi/\ln(R)>0$. The correspondence of two models of the ring is given by the explicit formula $z(u)=\exp(2\pi u/T)$.

In the strip model one can construct all real meromorphic differentials $d\xi$ invariant under the translations $u\to u+iT$.
The simplest of them is $d\xi=idu$, all the others are obtained via multiplication by a real (on the boundaries) meromorphic function $h(u)$ 
with due shift invariance. Reflection principle  meromorphically extends such a function to the whole plane where it has a period lattice 
$\mathbb{Z}+iT\mathbb{Z}$. Doubly periodic functions are effectively represented in terms of elliptic theta functions \cite{Ach}.
Choose a representative (modulo group of periods) $a_s^+$ of each zero of the function and a representative $a^-_s$ of its each pole. 
They satisfy a well-known lattice condition \cite{Ach} 
\be \label{Cond}
\sum_{s=1}^N (a_s^+-a_s^-)\in\mathbb{Z}+miT, 
\qquad m\in\mathbb{Z},
\qquad N=\deg h(u).
\ee
With this choice of representatives, the function is proportional to the following
\be
h(u)=\exp(-2\pi imu)\prod_{j=1}^N\frac{\theta_1(u-a_j^+)}{\theta_1(u-a_j^-)},
\ee
here $\theta_1(u)=\exp(-\pi T/4)\sin(\pi u)-\dots$ is the only odd theta function of the modulus $iT$.
Once the sets of zeros /poles are mirror symmetric with respect to the imaginary axis, one can derive 
it from the symmetries of theta function that the (suitably renormalized) function $h(u)$ is real on the boundaries of the strip:
$\{0,\frac12\}+i\mathbb{R}$. The differential  $d\xi=ih(u)du$ we transfer to the concentric ring by the exponential mapping $z(u)$,
lattice condition (\ref{Cond}) becomes the necessary and sufficient condition for the existing of a real differential 
with given sets of zeros and poles in a concentric ring:
\begin{enumerate}
\sf
\item The number of zeros equals the number of poles;
\item Total number of zeros and poles on each boundary (half-vortexes) is integer;
\item The sum of azimuths of poles is equal to the sum of azimuths of zeros modulo $\pi$.
\end{enumerate}
Here the azimuth is measured with respect to the center of the ring modulo $2\pi$ and zeros/poles on the boundary are counted with the
weight $\frac12$. 
\end{xmpl}

The authors of \cite{BM} have checked the validity of the last constraint for the positions of magnetic vortexes 
on the available experimental data and have found a good agreement with the experiment.

\subsection{Homologies and differentials on the Schottky double}
From the last example we've learned that  the positions of zeros/poles of a real differential in an annulus cannot be arbitrary.
Same effect holds for all domains of connectivity more than 1, essentially it stems from the Abel's theorem on the divisors of 
algebraic functions \cite{GrHar,Mu}. To explain this effect we introduce some auxiliary constructions.

\subsubsection{Even and odd basic cycles}
\begin{figure}[h!]
\includegraphics[scale=.85, trim = 1.5cm 22cm .5cm 1cm, clip]{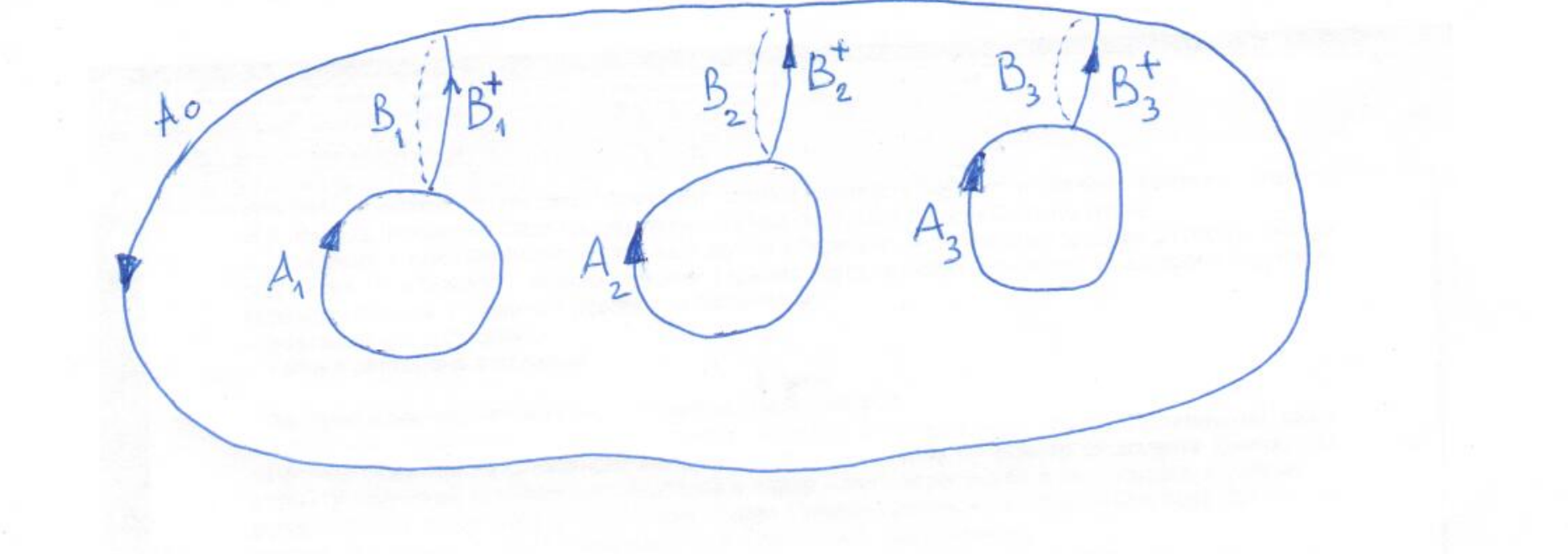}
\caption{Homology basis on the double of flat domain $\Omega$.} 
\label{CyclDbl}
\end{figure}

On a double ${\cal D}\Omega$ of $g+1$ connected domain $\Omega$ we introduce a basis $A_1,\dots, A_g$; $B_1,\dots, B_g$ 
of integer 1-homologies (=closed contours modulo certain equivalence) with the standard symplectic intersection form
\be
A_j\circ A_s= B_j\circ B_s=0, 
\quad A_j\circ B_s=\delta_{js}, 
\qquad j,s=1,\dots,g,
\ee
and additional mirror symmetry:
\be
\tau A_j=A_j;
\quad \tau B_j=-B_j, 
\qquad 
j=1,\dots,g.
\ee
The basis with the above properties is not unique. For instance, one can take the interior boundaries of $\Omega\subset{\cal D}\Omega$
with the induced orientation as $A$-cycles. We connect each of those to the exterior boundary $A_0\sim
-\sum_{j=1}^g A_j$ by $g$ mutually disjoint simple arcs $B_j^+$ and complete those to the odd cycles on the double:
$B_j:=B_j^+-\tau B_j^+$ as in Fig. \ref{CyclDbl}.

\subsubsection{Holomorphic differentials}
For representation of various function-theoretic objects on the Riemann surfaces they use e.g. holomorphic and meromorphic differentials.
Let us construct a basis of holomorphic differentials. There are $g$ so called harmonic measures $W_s(z)$ in the domain $\Omega$:
harmonic functions vanishing at all the boundaries but the unique interior one  $A_s$, where it is equal to $1$. Conjugate harmonic functions $H_s(z)$
are multivalued in the domain however the differentials $d\omega_s:=d(W_s+iH_s)$ are holomorphic and purely imaginary on the boundaries.
They may be holomorphically extended to the closed surface ${\cal D}\Omega$:
$$
\tau d\omega_s=-d\omega_s, \qquad s=1,\dots,g.
$$
The periods of those differentials are as follows:
$$
\int_{B_s}d\omega_j=-2\delta_{js}; 
\qquad
\int_{A_s}d\omega_j=i\Pi_{js}
$$
where $\Pi_{js}$ is a symmetric real matrix with positive diagonal, negative off-diagonal elements and diagonal dominance. 
It  may be interpreted as the capacity matrix. Indeed, $W_j$ may be considered as a voltage which induces the charge 
$\int_{A_s}(dW_j/dn)~dl=\int_{A_s}(dH_j/dl) ~dl=-i\int_{A_s}d\omega_j=\Pi_{js}$ on the boundary component $A_s$
($n$ and $l$ are normal and tangent directions to the boundary).

Real differentials $d{\bm\zeta}:=(d\zeta_1,\dots, d\zeta_g)^t $ with the usual A-normalization $\int_{A_s}d\zeta_j=\delta_{js}$ are related to the 
imaginary differentials $d {\bm\omega}:=(d\omega_1,\dots,d\omega_g)^t$ as
$$
i\Pi d\bm{\zeta}=d{\bm\omega}.
$$

\subsubsection{Meromorphic differentials}
\label{merodiff}
For the construction of magnetization patterns with singularities we need the differentials with simple poles.
Let $W_{pq}$ be a harmonic function in the domain $\Omega$ with just two logarithmic singularities of opposite signs at the points: $p$
from the closure of the domain $\Omega$ and $q$ from its interior, 
and homogenious Neumann boundary condition:
$$
\begin{array}{l}
W_{pq}(z)=m(p)\log|z-p|-\log|z-q|+ ~harmonic~ function\\
\partial W_{pq}(z)/\partial n~\vert_{z\in\partial\Omega}=0,
\end{array}
\qquad m(p)=\left\{
\begin{array}{ll}
1, & p\in Int~\Omega,\\
2, & p\in \partial\Omega.
\end{array}
\right.
$$ 
Again, let $H_{pq}(z)$ be a (multivalued) conjugate harmonic function in the domain, then the differential $d\eta_{pq}:=d(W_{pq}+iH_{pq})$
is single valued and real on the boundary of $\Omega$. It can be meromorphically extended to the closed surface ${\cal D}\Omega$
as 
$$
\tau d\eta_{pq}=d\eta_{pq},
$$
and satisfies $A-$ normalization conditions $\int_{A_s}d\eta_{pq}=0, \quad s=0,\dots,g$
(the integral is taken in the sense of principal value if $p$ lies on the boundary). 

\subsection{Constraints on the poles and zeroes}
Here we study the constraints on the positions of magnetization singularities in the multiconnected magnet
(described by a real differential) which are similar to that in the ring. 

The reflection $\tau$ naturally acts on the divisors -- finite formal sums of points $p\in{\cal D}\Omega$ with integer coefficients
$D=\sum_pm(p)\cdot p\to \tau D=\sum_pm(p)\cdot \tau p$. For any mirror symmetric divisor 
$D=\tau\cdot D$ -- for instance the divisor of zeros and poles of a real function or a differential -- 
we introduce its reduction $D/\tau$ as the divisor with the support in the closure of the domain $\Omega$, 
interior points inherit their multiplicity in $D$ and boundary points having half of that: 
$D$ equals to sum of the reduced divisor $D/\tau$ and 
its reflection. Each (reduced) divisor $D$ may be naturally decomposed into its positive and 
negative parts: $D=D^+-D^-$, $D^\pm\ge 0$.

\begin{thrm}
The reduced divisor of a real differential $d\xi$ in $\Omega$ is completely characterized as follows:
$(d\xi)/\tau=(d\omega)/\tau+D$ where $d\omega\neq0$ is any fixed real differential  
and $D=\sum_p m(p)\cdot p$, satisfies three conditions:
\be
\label{Deg0}
\deg D:=\sum_p m(p)=0;
\ee 
\be
\label{DegBd0}
\deg D~\vert_{A_s}:=\sum_{p\in A_s}m(p)\in\mathbb{Z}, \qquad s=0,\dots,g;
\ee
\be
\label{AbelCond2}
2\int_{D^+}^{D_-}d{\bf H}~\bigg(=2\sum_p m(p)\cdot{\bf H}(p)\bigg)\in\Pi\mathbb{Z}^g
\ee
where $d{\bf H}(z):=(dH_1,\dots,dH_g)^t$, 
and the integral with divisor limits is the sum of integrals between points of the same weight multiplied by this weight
(some interior points may be decomposed into halves if necessary). 
Real differential with the given reduced divisor is unique up to multiplication by a real constant and is given by the formula
\be
d\xi(z):=d\omega(z)\exp(\int_{z_0}^z\sum_p m(p)d\eta_{pq}),
\label{ReRepr} 
\ee
where $z_0\in A_0$, $q$ is any interior point of the domain $\Omega$ and $m(p)$ is the half-integer multiplicity of the point
$p$ in the reduced divisor $D$ satisfying (\ref{Deg0})-(\ref{AbelCond2}).
\end{thrm}
\begin{rmk}
Real conditions (\ref{AbelCond2}) do not depend on the choice of the auxiliary differential 
$d\omega$, the decomposition of the divisors in the limits of the integral into pairs of points and connecting them integration paths
within the domain $\Omega$.
\end{rmk}

\begin{xmpl}
Let us specify the constraints on the positions $D$ of (anti) vortexes in a pair of pants. Using the conformal invariance of real differentials, 
we conformally map arbitrary pants to the extended complex plane with three disjoint real slots removed as in Example \ref{Pants}. 
Now the double of the pants has the representation (\ref{PantsDble})  with the reflection $\tau(z,w)=(\bar{z}, -\bar{w})$. 
An auxiliary real holomorphic differential in this model has the appearance $d\omega=i\frac{(z-c)dz}w$ with real $c$.
We choose the symplectic basis of cycles as follows: $A_1,A_2$; $B_1,B_2$ are the inverse images of the segments 
$[e_1,e_2]$, $[e_5,e_6]$; $[e_2,e_3]$, $[e_4,e_5]$ under the projection $(z,w)\to z$, with suitable orientation.
The basis of imaginary holomorphic differentials has the appearance $d\omega_s=a_s\frac{(z-c_s)dz}w$
where real $a_s$ and $c_1\in [e_4,e_5]$, $c_2\in [e_2,e_3]$ may be found from 
the normalization conditions $\int_{B_s}d\omega_j=-2\delta_{js}$. Now the restrictions (\ref{AbelCond2}) for the 
positions of zeros $D^+$ and poles $D^-$ of a real differential read
$$
2Im\int_{D^+}^{D_-+c}d{\bm\omega}\in\Pi\mathbb{Z}^2.
$$
The latter condition does not depend on the real $c$: if $w^2(c)>0$, then the increment of $c$ 
results in the real increment of the latter integral; once $w^2(c)<0$, then $c$ represents a pair of half-points 
on the boundary of pants and the increment of $c$ does not change the total integral.
\end{xmpl}

{\bf Proof of the Theorem} \cite{SS,GrnL,FK} Let us fix the auxiliary real differential $d\omega$ say holomorphic $d\zeta_s$ or 
$id\omega_s$. The ratio of two real differentials is a real function: $h(z)=d\xi(z)/d\omega(z)$. Denote $D=\sum_pm(p)\cdot p$ 
the reduced divisor of  $h(z)$ and check three properties 
(\ref{Deg0})-(\ref{AbelCond2}). The first two are simple: $0=\deg(h)=2\deg D$; on each boundary component the (real) function 
changes its sign in even number of points $p$: zeros and poles of odd degree, exactly where $m(p)\in\frac12+\mathbb{Z}$.

Now we concentrate on (\ref{AbelCond2}). For any interior point $q\in\Omega$ the following representation is valid:
$$
dh/h=\sum_p m(p)d\eta_{pq},
$$
where $m(p)\in\mathbb{Z}/2$ is the multiplicity of the point $p$ in the reduced divisor of $h(z)$.
Indeed, both sides have the same singularities and the (principal parts of) integrals over each boundary component $A_s$ vanish.
Now the chain of equalities holds with ${\bf B}$ being the set of $B$-cycles on the surface:
\be
\label{chain}
2\pi i \mathbb{Z}^g\ni\int_{\bf B}\frac{dh}h=\sum_p m(p)\int_{\bf B}d\eta_{pq}\stackrel*=2\pi i\sum_p 2m(p)Re\int_q^pd{\bm\zeta}=
\ee
$$
2\pi i\sum_p 2m(p)Im~\Pi^{-1}\int_q^pd{\bf\omega}=2\pi i\sum_p 2m(p)~\Pi^{-1}\int_q^pd{\bf H}.
$$
Equality $*$ here is the Riemann bilinear relations (reciprocity law) and all the subsequent sums 
over $p$ are taken over the support of the reduced divisor $D$. We couple the terms with the opposite value of $m(p)$ in last sum
and get exactly the condition (\ref{AbelCond2}).

{\bf The inverse move.}  Suppose three conditions of the theorem hold for the reduced divisor $D$.
 One has to check three statements for the differential (\ref{ReRepr}):\\ 
(1) The differential $d\xi$ has the reduced divisor exactly $D+(d\omega)/\tau$ in the domain. Indeed, $\deg D=0$ and the integral in (\ref{ReRepr}) 
has no singularity at $z=q$.\\
(2) $d\xi$ is single valued in the domain. This follows from the $A$-normalization of elementary singular differentials 
$d\eta_{pq}$ and the condition (\ref{DegBd0}). \\
(3) The differential is real on the boundaries of the domain. Indeed, $2iIm\int_{B^+}d\eta_{pq}=\int_Bd\eta_{pq}$,
the sum of latter integrals over $p$ with the weights may be extracted from the chain of equalities (\ref{chain})
which together with the restriction (\ref{AbelCond2}) mean that the exponential factor in (\ref{ReRepr}) is real on the boundary of $\Omega$. 

\section{Prym real differentials and nonlocal merons}
D.J.Gross writes: 'Even though we have found an infinite class of solutions of the nonlinear equations of motions, it is clear that we do 
not have the most general solution..'\cite{Gr}.
All the ommited meronic solutions will be listed in this section. Let us recall (Sect.\ref{Sect1}) that the general local solution of 
(\ref{ELeq}) with $|w(z)|=1$ away from the singularities of the magnetization has the appearance $w(z)=f/|f|$ with holomorphic 
$f(z)\neq 0$. This function may be not single valued globally and yet bring us to the single valued magnetization iff $f(z)$ is multiplied 
by a positive function once we circumflex the singularities and/or boundary components of the magnet. This positive function necessarily 
reduces to a constant which gives us the monodromy:
\be
\label{Mndrmy}
\rho:\quad
\pi_1(\Omega-Supp(f))\to\mathbb{R}^+.
\ee
The (commutative) monodromy in a finitely punctured domain is completely determined by the monodromy $\rho(p)$ of a loop encompassing a puncture 
$p\in Supp(f)$ counterclockwise and the monodromies $\rho(A_s)$ of oriented boundary components with the only relation 
$$
\prod_p\rho(p)=\prod_{s=0}^g\rho(A_s).
$$

Let us specify the appearance of $f(z)$ near an isolated singularity $z=p$ of magnetization. 
The function $h(z):=(z-p)^{-i\mu}$, $2\pi\mu=\log(\rho(p))$, has the same local monodromy as $f$ has and therefore the function $f/h$
is single valued around $z=p$. This point cannot be an essential singularity of $f/h$, otherwise the lose the property of energy log divergence 
-- see Sect. \ref{Sect1}, but it can be a pole or a zero. Eventually we have the local behavior of the multivalued differential 
$d\xi:=dz/f(z)$ near its divisor: 
\be
\label{PrymLoc}
d\xi(z)=(z-p)^{m+i\mu}\phi(z)dz,
\qquad\phi(z)\in{\cal O}^*(p), \quad m\in\mathbb{Z}, \quad\mu\in\mathbb{R},
\ee
with holomorphic function $\phi(z)$ nonvanishing in the vicinity of $z=p$. This differential has positive monodromy and 
as before it is real on the boundary of the domain. Latter property is not spoiled by the monodromy. 
Such objects are called real Prym differentials \cite{FK,Chue}.

The divisor $D=(d\xi)$ of a real Prym differential $d\xi$ has complex weights $c(p)$ with (integer in our case) real parts determined by the 
growth of the differential near the singularity $z=p$ and imaginary part related to positive local monodromy.  
This divisor remains intact under the natural action of the reflection $\tau$ on complex valued divisors
$$
\quad D:=\sum_p c(p)\cdot p \to  
\tau D:=\sum_p \overline{c(p)}\cdot\tau p,
$$
in particular points on the real ovals have purely real weight. For the symmetric complex valued divisor $D=\tau\cdot D$ 
we introduce its reduction $D/\tau$ to the domain $\Omega$ as a divisor with the support in the closure of the domain,
multiplicities of interior points intact and multiplicities of boundary points divided by 2.

\begin{thrm}
The reduced divisor of a real Prym differential $d\xi$ in $\Omega$ with the given monodromy
$\rho:~\pi_1(\Omega- Supp~(d\xi))\to\mathbb{R}^+$
is completely characterized as follows:
$(d\xi)/\tau=(d\omega)/\tau+D$ where $d\omega\neq0$ is any fixed real differential  
and the reduced complex valued divisor $D=\sum_p c(p)\cdot p$, satisfies four conditions:\\
 \begin{enumerate}
 \item Real part of $D+\tau D$ is integer and has zero degree.
 \item Imaginary part of $D$ is determined by the monodromy: $Im~D=2\pi\sum_p\log\rho(p)\cdot p$.
 \item $\deg Re~D~\vert_{A_s}:=\sum_{p\in A_s}Re~c(p)\in\mathbb{Z}, \qquad s=0,\dots,g;$
 \item The lattice condition holds:
 \be
 2Im~\sum_p c(p)\int_{z_0}^pd{\bm\omega} +\frac1\pi\log\rho({\bf A})\in
 \quad\Pi\mathbb{Z}^g. 
 \label{AbelCond3}
 \ee
 here $\log\rho({\bf A})$ is the vector $(\log\rho(A_1),\dots,\log\rho(A_g))^t$, $z_0\in A_0$.
 \end{enumerate}
 
 Real Prym differential is reconstructed from its divisor uniquely up to multiplication by a real constant:
 \be
 \label{Prym}
 d\xi(z)=d\omega(z)\exp(\int_{z_0}^zd\eta+\sum_{s=1}^g\log\rho(A_s)d\zeta_s)
 \ee
 where $d\eta$ is (real single valued)  $A$-normalized differential on the double ${\cal D}\Omega$
 with simple poles only and the residue divisor $D+\tau D$.
\end{thrm}

{\bf Proof} of this theorem more or less follows the arguments from Theorem 1. The ratio $h(z)=d\xi/d\omega$
is a real Prym function with the same monodromy as $d\xi$ and the reduced divisor $D$. 
Real single valued differential $d\log h(z)$, may be decomposed as
$$
d\log h = d\eta+d\zeta
$$
with the $A$-normalized differerential $d\eta$ having the same singularities as $d\log h$ 
(now it cannot be decomposed into the elementary terms $d\eta_{pq}$ from Sect.\ref{merodiff}; we use the existence theorem \cite{FK}) 
and the real holomorphic differential $d\zeta$ whose further decomposition into the basic differentials $d\zeta_s$ may be found by 
integration of the both sides of the latter equality along the real ovals. The inclusion 
$$
2\pi i\mathbb{Z}^g\ni 2Im \int_{{\bf B}^+}d\log h=\int_{\bf B}d\log h
$$
after the substitution of the above decomposition and the usage of Riemann bilinear relations we eventually get the lattice condition 
(\ref{AbelCond3}) of the theorem.

\subsection{Examples}
\subsubsection{Disc} 
We list all non-local meron solutions in a simply connected domain. Such domains may be conformally mapped to a half-plane and we assume that 
$\Omega=\mathbb{H}$. The general real Prym differential takes the form $d\xi(z)=h(z)dz$ with real functions $h(z)$
being the product of several elementary functions of the kind
$$
h^\pm(z|a,\mu)=
(z-a)^{\pm1+i\mu}(z-\bar{a})^{\pm1-i\mu},
\qquad a\in\mathbb{H}, \mu\in\mathbb{R},
$$
$$
h^\pm(z|c)=(z-c)^{\pm1}, 
\qquad c\in\mathbb{R}.
$$
with various parameters $\pm,~a,~\mu,~c$.

\subsubsection{Annulus} 
Here we evaluate in terms of elliptic theta functions all the non-local meron solutions in the annulus represented as the factor  
of a strip $0<Re~u<\frac12$ by translations, the model  introduced in the last example of Sect. \ref{Examples}.
The map to a concentric ring is given by the exponential function $z(u)=\exp(2\pi u/T)$, $T>0$. 
General solution $d\xi(u)=ih(u)du$ may be decomposed into the elementary ones: 1) without any singularities (holomorphic), 
2) two interior vortexes and 3) one interior vortex and two half vortexes on the boundary. Three mentioned cases
correspond to the choice of the real meromorphic function $h(u)$ in the strip as follows.

\begin{enumerate}
  \item $$h(u|m)=\exp(2\pi imu), \qquad m\in\mathbb{Z}.$$
The integral lines of the magnetization field $w(z)$ make up $|m|$ embedded Rieb foliations in the concentric annulus 
with interior magnetic domain walls -- the concentric limit cycles of the 
magnetization field. The patterns of this type were observed in the experiments.
\item $$
h(u|a_\pm,\mu_\pm)=\frac{\theta_1^{1+i\mu_+}(u-a_+)\theta_1^{1-i\mu_+}(u+\overline{a_+})}
{\theta_1^{1+i\mu_-}(u-a_-)\theta_1^{1-i\mu_-}(u+\overline{a_-})},
\qquad\mu_\pm\in\mathbb{R};
\quad 0<Re~ a_\pm<1/2.
$$
This solution corresponds to a vortex and antivortex in two interior points
of the ring with the reduced divisor $D=(1+i\mu_+)\cdot z(a_+)+(-1+i\mu_-)\cdot z(a_-)$
which determines the monodromy.
\item
$$
h(u|a,a_1,a_2,\mu)=\frac{\theta^{1+i\mu}_1(u-a)\theta^{1-i\mu}_1(u+\overline{a})}
{\theta_1(u-a_1)\theta_1(u-a_2)},
\qquad\mu\in\mathbb{R};
\quad Re a_1=Re a_2 \in\{0,\frac12\}; 0<Re a<\frac12 
$$
The magnetization pattern contains a vortex in the interior of the annulus and two half-vortexes on its boundary component.  
The reduced divisor of the real Prym differential $d\xi=ih(u)du$ transferred to the annulus is $(1+i\mu)\cdot z(a)-\frac12\cdot z(a_1)-
\frac12\cdot x(a_2)$.
\end{enumerate}
In all three cases one checks that the function $h(u)$ is $1-$periodic, real on the strip boundaries
and acquires a constant positive factor under the transformation $u\to u+iT$. 

\section{Conclusion} 
This paper is devoted to the mathematical aspects of the description of 
topologically charged magnetization textures \cite{Met1, Metlov} in multiply-connected planar nanomagnets.
New multi vortex configurations -- the exact solutions of the nonlinear boundary value problem 
which describe the micromagnetic patterns -- were introduced in \cite{BM1} in terms of the so called 
real meromorphic differentials on the closed Riemann surface -- 
the Schottky double of the original domain. This approach highlights the conformal invariance of the problem and allows to give
effective formulas for the solutions in algebraic terms, in terms of theta functions \cite{Ach, Mu}, Schottky functions \cite{B} and 
Schottky-Klein Prime form \cite{Mu, B1, BM1}. 
It permitted us to list explicitly all the magnetization textures, following from the model \cite{Met1, Metlov}, 
in  two- and  three-connected domains. We established the complete set of constraints on the numbers and the positions of the 
singularities of magnetization for arbitrary multiconnected domain.
This extends our earlier result on the constraints on positions of topological singularities 
in an annulus \cite{BM} which were checked in the experiment.
The constraints on the positions of magnetization singularities may be of interest in spintronic application:
once the vortex patterns bear the information we have a means to check if there were errors in reading the information off.

Most interestingly, the present representation for the meronic
magnetization field allows to exploit its invariance with respect to
rescaling of the underlying meromorphic function. Mathematically this
leads to non-local objects -- Prym differentials on the double.
Those solutions have new degrees of freedom -- the local and global monodromy -- which may serve as Ritz parameters for 
the next in the magnitude terms of Hamiltonian  or the dynamical variables for the adiabatic approach.

\vspace{5mm}
\parbox{7cm}
{\it
Moscow Inst. of Physics and Technology;\\
Moscow State University;\\
Institute for Numerical Mathematics,\\
Russian Academy of Sciences;\\[1mm]

{\tt ab.bogatyrev@gmail.com}}


\begin{thebibliography}{15}
\bibitem{Ach} Achiezer N.I.,  ~Elements of Theory of Elliptic Functions -- Nauka, Moscow, 1970 
\bibitem{Aha} A. Aharoni, Introduction to the theory of ferromagnetism  --
Oxford University Press, Oxford, 1996
\bibitem{GrnL} N.L. Alling, N.Greenleaf ~Foundations of the Theory of Klein Surfaces --
Springer Berlin Heidelberg, 1971 
\bibitem{BP} Belavin A.A., Polyakov A.M.//JETP Letters, 1975
\bibitem{B} A.Bogatyrev, Computations in moduli spaces// Computational Methods and Function Theory, 7 (2007), No. 2, 309-324.
\bibitem{B1} A.Bogatyrev, Prime form and Schottky model// Computational Methods and Function Theory, 9:1 (2009), 47-55.
\bibitem{B2} Bogatyrev A. Elementary construction of Jenkins-Strebel differentials.// Math.Notices 2012, 91:1, pp. 143--146.
\bibitem{BM1} A.B. Bogatyrev, K.L. Metlov 
~Magnetic states in multiply-connected flat nano-elements//Low temperature physics, 51:10 (2015) pp. 984-988., arXiv:1504.01162
\bibitem{BM} Bogatyrev A.B., Metlov K.L. ~Topological constraints on positions of magnetic solitons in 
multiply-connected planar magnetic nano-elements //submitted to PRB, arXiv:1609.02509
\bibitem{Chue} Chueshov V.V. Multiplicative functions and Prym differentials on a variable Riemann surface (Russian) -- Kemerovo U. Publishing, 2003.
\bibitem{FK} H.M. Farkas and I.Kra,  ~Riemann Surfaces
-- Springer Verlag, NY, Heidelberg, Berlin, 1980
\bibitem{GrHar}  Ph.Griffiths J.Harris, ~Principles of Algebraic Geometry, vol. 1 and 2 -- Wiley,  1994.
\bibitem{Gr} D.J.Gross, ~Meron configurations in the two-dimensional  O(3) $\sigma$-model
// Nucl.Phys.B 132(1978), 439-456.
\bibitem{Met1} K. L. Metlov, Two-dimensional topological solitons in soft ferromagnetic cylinders.// arXiv:cond-mat/0102311
\bibitem{Metlov} K. L. Metlov, Magnetization patterns in ferromagnetic nano-elements as functions of complex
variable// Phys. Rev. Lett. 105, 107201 (2010).
\bibitem{Mu} D.Mumford, ~Tata lectures on theta I,II,III -- Birkha\"user, 1983.
\bibitem{SS} M.Schiffer, D.C. Spencer ~Functionals of Finite Riemann Surfaces --
Princeton University Press, 1954
\end{thebibliography}
\end{document}